# Evaluating Investment Performance: The p-index and Empirical Efficient Frontier

Jing Li [a, *], Bowei Guo [a], Xinqi Xie [a], Kuo-Ping Chang [a, **]

August 2025

[a] Jinhe Center for Economic Research, Xi'an Jiaotong University, Xianning West Road, 28#, Xi'an, Shaanxi Province, 710049, China.

* Corresponding author.

  E-mail addresses: jing.li.jcer@xjtu.edu.cn

** E-mail addresses: kpchang@mx.nthu.edu.tw

Evaluating Investment Performance: The p-index and Empirical Efficient Frontier

## ABSTRACT

This paper has used European put option to construct the p-index to measure the underlying asset's risk level. The p-index measures the insurance fee for each insured dollar to guarantee that the asset achieves at least a $\delta$ rate of return on a specified future date. The study also uses the p-index to construct both the p-ratio and the empirical efficient frontier (EEF) to examine the performance of different investment strategies for stocks listed on both China's SSE Composite Index and the US NYSE Composite Index from 2014 to 2024. The empirical results have shown that firstly, with one-week holding period and reinvesting, for SSE Composite Index stocks, the highest p-ratio investment strategy produces the largest annualized rate of return, at 499.97%; and for NYSE Composite Index stocks, all the three strategies with both one-week and one-month periods generate negative returns. Secondly, with non-reinvesting, for SSE Composite Index stocks, the highest p-ratio strategy with one-week holding period yields the largest annualized rate of return (36.48%); and for NYSE Composite stocks, the one-week EEF strategy produces a 10.44% annualized return. Thirdly, under the one-week EEF investment strategy, for NYSE Composite Index stocks, the right frontier yields a higher annualized return, but for SSE Composite Index stocks, the left frontier (stocks on the empirical efficient frontier) yields a higher annualized return than the right frontier. Fourthly, for NYSE Composite Index stocks, there is a positive linear relationship between monthly return and the p-index, but no such relationship is evident for SSE Composite Index stocks. Fifthly, for NYSE Composite Index stocks, with the reinvesting strategy, the highest p-index portfolio (H) yields the largest annualized return (20.58%) and the H-L portfolio (long H, short L) generates an annualized return of 22.85%. For SSE Composite Index stocks, the lowest p-index portfolio (L) yields the largest annualized return (27.36%). Sixthly, for NYSE Composite Index stocks, the traditional five-factor model performs poorly, and adding the p-index as a sixth factor provides incremental information. In contrast, for SSE Composite Index stocks, adding the p-index may not provide such improvement.

# 1. Introduction

Risk level can be defined as the probability that a party will fail to meet a stated obligation or promise. In finance literature, risk is commonly quantified using metrics such as beta (derived from the Capital Asset Pricing Model, CAPM) or the variance of an asset's rate of return. Markowitz's (1952) mean-variance portfolio analysis introduces the efficient frontier (the set of optimal portfolios offering the highest expected return for a specific level of risk, measured by the standard deviation of return) for investment. Statman and Clark (2013) offered the efficient range, the location of portfolios that acknowledge imprecise estimates of mean-variance parameters and accommodate investor preferences beyond high mean and low variance, as a replacement for the efficient frontier. Cong et al. (2015) construct a series of test assets using a new class of tree-based models, which significantly advances the efficient frontier, resulting in factor models outperforming popular ones for investment and cross-sectional pricing.

Fama and MacBeth (1973) used a two-step regression method to estimate risk premiums and evaluate asset pricing models. Although theoretically appealing, the CAPM has demonstrated unsatisfactory empirical performance. A series of studies have identified additional factors that explain asset returns. Banz (1981) and Basu (1983) documented the significant roles of firm size and the earnings-price ratio, respectively. This line of inquiry culminated in the Fama-French three-factor model, which incorporated market risk, size, and value factors to better capture return variations (Fama and French, 1993). Their work showed that the CAPM fails to account for all sources of stock returns, with Fama and French (1996) finding no positive relationship between average returns and market betas.

Further research expanded the model to include other influences. The momentum effect, identified by Jegadeesh and Titman (1993), was integrated by Carhart (1997) into a four-factor model to explain mutual fund performance. Subsequent studies showed that its profitability is contingent on market state (Cooper, Gutierrez, and Hameed, 2004) and that it coexists with long-term return reversals and contrarian strategies (DeBondt and Thaler, 1985; Conrad and Kaul, 1998). The expanding framework also incorporated other factors, such as the aggregate liquidity premium identified by Pástor and Stambaugh (2003) and the profitability and investment factors added by Fama and French (2015) to their 1993 model of market, size, and value risks. However, even these powerful multi-factor models face significant challenges. First, the economic interpretation of factors remains contentious—for instance, whether the value factor (HML) represents a risk premium or behavioral bias. Second, factor premiums often exhibit time decay (Green et al., 2017), suggesting that factors based on traditional financial indicators may be prone to overfitting, casting doubt on their long-term robustness.

This paper has used European put option to construct the p-index to measure the underlying asset's risk level (Chang, 2021 and 2023). The p-index measures the insurance fee for each insured dollar to guarantee that the asset achieves at least a $\delta$ rate of return on a specified future date. The study also uses the p-index to construct both the p-ratio and the empirical efficient frontier (EEF) to examine the performance of different investment strategies for stocks listed on both China's SSE Composite Index and the US NYSE Composite Index from 2014 to 2024. The empirical results have shown that firstly, with one-week holding period and reinvesting, for SSE Composite Index stocks, the highest p-ratio strategy produces the highest annualized rate of return, at 499.97%; and for NYSE Composite Index stocks, all the three strategies with both one-week and one-month holding periods generate negative returns. Secondly, with non-reinvesting, for SSE Composite Index stocks, the highest p-ratio strategy with one-week holding period yields the highest annualized rate of return (36.48%); and for NYSE Composite stocks, the one-week EEF strategy produces a 10.44% annualized return. Thirdly, for NYSE Composite Index stocks, a positive linear relationship exists between monthly return and the p-index, whereas no such relationship is evident for SSE Composite Index stocks. Also, for NYSE Composite Index stocks, with the reinvesting strategy, the highest p-index portfolio (H) yields the largest annualized return (20.58%) and the H-L portfolio (long H, short L) generates an annualized return of 22.85%. For SSE Composite Index stocks, the lowest p-index portfolio (L) yields the largest annualized return (27.36%). Fourthly, for NYSE Composite Index stocks, the traditional five-factor model performs poorly, and adding the p-index as a sixth factor provides incremental information. In contrast, for SSE Composite Index stocks, adding the p-index may not provide such improvement.

The remainder of this paper is organized as follows. Section 2 derives the p-index and its upper and lower bounds, and the empirical efficient frontier. It also discusses the properties of the p-index under the binomial asset pricing model. Data and empirical results are presented in Section 3. Concluding remarks appear in Section 4.

## 2. The p-index and Empirical Efficient Frontier

**The p-index as a Measure for Risk Level**

Assume that an asset's current price at $t = 0$ is $S_0$, and its value at $t = T$ is an uncertain $S_T$. For this asset, at $t = 0$ buying a European call option $c$ with the strike price $K$ will give the owner of the call: $Max[S_T - K, 0]$ at $t = T$. At $t = 0$ buying a European put option $p$ with the strike price $K$ will give the owner of the put: $Max[K - S_T, 0]$ at $t = T$. Thus, both $p$ and $c$ can be interpreted as insurance. The put option $p$ can be interpreted as: If at $t = 0$ a person buys both the underlying asset $S_0$ and an insurance $p$, then at $t = T$ the value of her owning the underlying asset will be worth at least $K$, i.e., $Max[S_T, K]$. That is, by paying $p$ as an insurance fee to an insurance company at $t = 0$, the insurance company will ensure that the underlying asset can deliver at least $K$ at $t = T$: $Max[S_T, K]$. Higher $p$ means less likelihood that the underlying asset will be worth at least $K$. The call option $c$, on the other hand, can be interpreted as: If at $t = 0$ a person short-sells the underlying asset $S_0$ and buys an insurance $c$ (where $K$ can be interpreted as the insurance deductible), then at $t = T$ this person will not need to pay more than $K$ to buy back the asset. That is, at $t = T$, to buy back the underlying asset this person will pay $Min[S_T, K]$ and the insurance company will pay $Max[S_T - K, 0]$ such that $S_T = Min[S_T, K] + Max[S_T - K, 0]$.

For any asset which delivers an uncertain payoff $S_T$ at $t = T$, there exists a corresponding put-call parity at $t = 0$:[1]

$$c + \frac{K}{1+r} = S_0 + p. \tag{1}$$

In eq. (1), if $K > (<)(1+r)S_0$, then $p > (<)c$. If $K = (1+r)S_0$, then $p = c$. That is, for the insurance company, when $K = S_0(1+r)$, the risk of insuring that the underlying asset is worth at least $K$ at $t = T$ is equivalent to the risk of insuring that the insurant needs not to pay more than $K$ to buy the asset at $t = T$.[2]

---

[1] Consider two portfolios at $t = 0$ and a simple risk-free interest rate $r$:

Portfolio A: one European call option $c$ with strike price $K$, and cash $\frac{K}{1+r}$ deposited in a bank;

Portfolio B: one European put option $p$ with strike price $K$, and one unit of the underlying asset $S_0$.

On the expiration date $t = T$, both portfolios give exactly the same payoff: $Max[S_T, K]$. Thus, the costs of the two portfolios at $t = 0$ must be the same, i.e., $c + \frac{K}{1+r} = S_0 + p$.

[2] We can rewrite eq. (1) as: $S_0 = c + \left(\frac{K}{1+r} - p\right)$. If $S_0$ is the market value of a levered firm, then $c$ is the equity, $\left(\frac{K}{1+r} - p\right)$ is the risky debt, and $p$ is the insurance to ensure the promised payment $K$ to debtholders at $t = T$. Note that the changes of $K$ will not affect $S_0$, i.e., the Modigliani-Miller capital structure irrelevancy proposition is an example of 'financial diversification irrelevancy', see Chang (2023). With $K > 0$ and $S_0 > 0$, the risky debt must be strictly positive, i.e.,

Suppose that there are $n$ uncertain assets at $t = 0$. We can rewrite eq. (1) as:

$$S_{0i} = c_i + \frac{K_i}{1+r} - p_i, \quad i = 1, \dots, n, \tag{2}$$

and divide both sides of the above equation by $S_{0i}$:

$$1 = \frac{c_i}{S_{0i}} + \frac{K_i/S_{0i}}{1+r} - \frac{p_i}{S_{0i}}, \quad i = 1, \dots, n. \tag{3}$$

Let $K_i = S_{0i}(1+\delta)$ where $\delta > -1$, we have:

$$1 = \frac{c_i}{S_{0i}} + \frac{1+\delta}{1+r} - \frac{p_i}{S_{0i}}, \quad i = 1, \dots, n. \tag{4}$$

That is, for the i-th one-dollar asset (e.g., i-th one-dollar stock) at $t = 0$, to ensure that at $t = T$, this asset will deliver at least a $\delta$ rate of return (i.e., one-dollar becomes at least $1 + \delta$ dollars), the insurance fee paid at $t = 0$ is: $\frac{p_i}{S_{0i}}$. Hence, if $\frac{p_i}{S_{0i}} > \frac{p_j}{S_{0j}}$, it means that the risk level of the i-th asset delivering at least a $\delta$ rate of return is greater than that of the j-th asset.

Divide both sides of eq. (4) by $1 + \delta$:

$$\frac{1}{1+\delta} = \frac{c_i}{S_{0i}(1+\delta)} + \frac{1}{1+r} - \frac{p_i}{S_{0i}(1+\delta)}, \quad i = 1, \dots, n. \tag{5}$$

We can define the p-index of the i-th asset as:

$$\text{p-index:} \quad \frac{p_i}{K_i} = \frac{p_i}{S_{0i}(1+\delta)}, \quad i = 1, \dots, n. \tag{6}$$

The p-index measures the insurance fee for each insured dollar to guarantee that the asset achieves at least a $\delta$ rate of return at $t = T$. Thus, higher p-index means higher risk (i.e., less likelihood) for the asset $S_0$

---

$\left(\frac{K}{1+r} - p\right) > 0$, or $\frac{K}{1+r} > p$ and $\frac{K}{1+r} \neq p$.

to deliver at least a $\delta$ rate of return at $t = T$. As shown by Chang (2020, 2023), upper and lower bounds for the put option $p$ is: $Max\left[\frac{K}{1+r} - S_0, 0\right] \leq p < \frac{K}{1+r}$, i.e., an asset cannot sell for more than or equal to the present value of a sure payment of its maximum payoff. Hence, upper and lower bounds for the p-index is: $Max\left[\frac{1}{1+r} - \frac{1}{1+\delta}, 0\right] \leq \frac{p}{S_0(1+\delta)} < \frac{1}{1+r}$, where $K = S_0(1 + \delta)$.

After deriving the p-index (i.e., risk level) of an asset, we can also calculate the p-ratio of the asset:

$$\text{p-ratio:} \quad \frac{R_i - r}{\text{p-index}} = \frac{R_i - r}{p_i/[S_{0i}(1+\delta)]}, \quad i = 1, \dots, n, \tag{7}$$

where $R_i$ is the rate of return of the i-th asset, and $r$ is the simple risk-free interest rate. The p-ratio, as the Treynor ratio, can be defined as an asset's risk-adjusted excess return, i.e., how much return an asset investment earned for the amount of risk the asset investment assumed.

Suppose that in eq. (4), investing $w_i$ dollars in one-dollar stock $i$, where $w_i \geq 0$ and $\sum_{i=1}^{n} w_i = 1$:

$$w_i = (w_i)\frac{c_i}{S_{0i}} + (w_i)\frac{1+\delta}{1+r} - (w_i)\frac{p_i}{S_{0i}}, \quad i = 1, \dots, n. \tag{8}$$

Summing all $n$ equations (for $i = 1, \dots, n$) is equivalent to investing $\$1$ to form a portfolio with weights $w_i \geq 0$ and $\sum_{i=1}^{n} w_i = 1$:

$$1 = \sum_{i=1}^{n} w_i = \sum_{i=1}^{n} \frac{w_i c_i}{S_{0i}} + \frac{1+\delta}{1+r} - \sum_{i=1}^{n} \frac{w_i p_i}{S_{0i}}. \tag{9}$$

That is, to ensure that at $t = T$, this one-dollar investment in the portfolio will deliver at least a $\delta$ rate of return (i.e., one dollar becomes at least $1 + \delta$ dollars), the insurance fee paid at $t = 0$ is: $\sum_{i=1}^{n} \frac{w_i p_i}{S_{0i}}$. By dividing both sides of eq. (9) by $1 + \delta$, we have:

$$\frac{1}{1+\delta} = \frac{1}{1+\delta}\sum_{i=1}^{n} \frac{w_i c_i}{S_{0i}} + \frac{1}{1+r} - \frac{1}{1+\delta}\sum_{i=1}^{n} \frac{w_i p_i}{S_{0i}}. \tag{10}$$

Thus, the p-index of this portfolio is:

$$\frac{1}{1+\delta}\sum_{i=1}^{n}\frac{w_i p_i}{S_{0i}} = \sum_{i=1}^{n}\left[w_i \cdot \frac{p_i}{S_{0i}(1+\delta)}\right], \tag{11}$$

and the p-index of the equally-weighted portfolio (where $w_i = 1/n$, $i = 1, \ldots, n$) is:

$$\frac{1}{n(1+\delta)}\sum_{i=1}^{n}\frac{p_i}{S_{0i}} = \frac{1}{n}\sum_{i=1}^{n}\frac{p_i}{S_{0i}(1+\delta)}. \tag{12}$$

Also, from eq. (2), summing all $n$ equations (where $K_i = S_{0i}(1+\delta), i = 1, \ldots, n$), we have the value-weighted portfolio:

$$\sum_{i=1}^{n}S_{0i} = \sum_{i=1}^{n}c_i + \frac{(1+\delta)\sum_{i=1}^{n}S_{0i}}{1+r} - \sum_{i=1}^{n}p_i. \tag{13}$$

By dividing both sides of eq. (13) by $\sum_{i=1}^{n}S_{0i}$, we have:

$$1 = \frac{\sum_{i=1}^{n}c_i}{\sum_{i=1}^{n}S_{0i}} + \frac{(1+\delta)}{1+r} - \frac{\sum_{i=1}^{n}p_i}{\sum_{i=1}^{n}S_{0i}} \tag{14}$$

or

$$\frac{1}{1+\delta} = \frac{1}{1+\delta}\cdot\frac{\sum_{i=1}^{n}c_i}{\sum_{i=1}^{n}S_{0i}} + \frac{1}{1+r} - \frac{1}{(1+\delta)}\cdot\frac{\sum_{i=1}^{n}p_i}{\sum_{i=1}^{n}S_{0i}}. \tag{15}$$

Thus, the p-index of the value-weighted portfolio is:

$$\frac{1}{(1+\delta)}\cdot\frac{\sum_{i=1}^{n}p_i}{\sum_{i=1}^{n}S_{0i}} = \frac{\sum_{i=1}^{n}p_i}{\sum_{i=1}^{n}S_{0i}(1+\delta)}. \tag{16}$$

**The Empirical Efficient Frontier**

Let $R_i$ denote the i-th stock's rate of return and let $v_i$ denote the i-th stock's p-index: $\frac{p_i}{S_{0i}(1+\delta)}$. From eq. (11), the p-index of the portfolio (where $w_i \geq 0$ and $\sum_{i=1}^{n}w_i = 1$) is: $\sum_{i=1}^{n}w_i v_i$. By varying $\bar{R}$ in $\sum_{i=1}^{n}w_i R_i = \bar{R}$, we can use the following linear programming model to draw the curve of minimum p-index portfolios:

$$\text{Minimize} \quad v = w_1 v_1 + w_2 v_2 + w_3 v_3 + \cdots + w_n v_n$$
$$\text{Subject to}$$
$$w_1 R_1 + w_2 R_2 + w_3 R_3 + \cdots + w_n R_n = \bar{R}$$
$$\sum_{i=1}^{n} w_i = 1$$
$$w_i \geq 0, i = 1, \ldots, n. \tag{17}$$

Also, by varying $\bar{v}$ in $\sum_{i=1}^{n} w_i v_i = \bar{v}$, we can use the following linear programming model to draw the curve of maximum rate of return portfolios:

$$\text{Maximize} \quad R = w_1 R_1 + w_2 R_2 + w_3 R_3 + \cdots + w_n R_n$$
$$\text{Subject to}$$
$$w_1 v_1 + w_2 v_2 + w_3 v_3 + \cdots + w_n v_n = \bar{v}$$
$$\sum_{i=1}^{n} w_i = 1$$
$$w_i \geq 0, i = 1, \ldots, n. \tag{18}$$

The intersection of the set of minimum p-index portfolios from eq. (17) and the set of maximum rate of return portfolios from eq. (18) is the empirical efficient frontier (EEF); see an example in Figure 1 in Section 3.

**The Binomial Case**

The binomial option pricing model may be presented as the follows.

$$c_u = Max[S_0 \cdot u - K, 0]$$
$$p_u = Max[K - S_0 \cdot u, 0]$$

$$S_0 \xrightarrow{\pi} S_0 \cdot u$$
$$S_0 \xrightarrow{1-\pi} S_0 \cdot d$$

$$c = ? \qquad c_d = Max[S_0 \cdot d - K, 0]$$

$$p = ? \qquad\qquad p_d = Max[K - S_0 \cdot d, 0]$$

where $u > 1 + r$, $0 < d < 1 + r$, $r$ is the simple risk-free interest rate, and $K$ is the strike price.

From the Gordan theory we have:[3]

$$\begin{cases} \text{Money market:} \quad 1 = \frac{1}{1+r}[\pi(1+r) + (1-\pi)(1+r)] \\ \text{The asset:} \quad S_0 = \frac{1}{1+r}[\pi \cdot S_0 u + (1-\pi) \cdot S_0 d] \\ \text{Call option:} \quad c = \frac{1}{1+r}[\pi \cdot Max(S_0 u - K, 0) + (1-\pi) \cdot Max(S_0 d - K, 0)] \\ \text{Put option:} \quad p = \frac{1}{1+r}[\pi \cdot Max(K - S_0 u, 0) + (1-\pi) \cdot Max(K - S_0 d, 0)] \end{cases} \quad (19)$$

where $S_0 d < K < S_0 u$, $\pi = \frac{(1+r)-d}{u-d}$ and $1 - \pi = \frac{u-(1+r)}{u-d}$. Thus, let $K = S_0(1 + \delta)$, the p-index for the asset $S_0$ is:

$$\frac{p}{K} = \frac{p}{S_0(1+\delta)} = \frac{\frac{1-\pi}{1+r}[S_0(1+\delta) - S_0 d]}{S_0(1+\delta)} = \frac{(1+\delta)-d}{1+\delta} \cdot \frac{1-\pi}{1+r}. \qquad (20)$$

It shows that if at $t = 0$ an asset $S_0$ has a lower down move $d$, it implies higher risk (i.e., a lower likelihood) for the asset to achieve at least a $\delta$ rate of return at $t = T$, i.e., $\frac{\partial}{\partial d}\left[\frac{p}{(1+\delta)S_0}\right] = \frac{-1}{1+\delta} \cdot \frac{1-\pi}{1+r} < 0$, $\frac{\partial^2}{\partial d^2}\left[\frac{p}{(1+\delta)S_0}\right] = 0$; and $\frac{\partial}{\partial \delta}\left[\frac{p}{(1+\delta)S_0}\right] = \frac{d}{(1+\delta)^2} \cdot \frac{1-\pi}{1+r} > 0$, $\frac{\partial^2}{\partial \delta^2}\left[\frac{p}{(1+\delta)S_0}\right] < 0$. Also, assets having $d = 0$ will have exactly the same constant p-index: $\frac{p}{K} = \frac{p}{S_0(1+\delta)} = \frac{1-\pi}{1+r}$.

---

[3] Chang (2015, p. 41) has shown the Gordan theory:

Let $A$ be an $m \times n$ matrix. Then, exactly one of the following systems has a solution:

System 1: $Ax > 0$ for some $x \in R^n$

System 2: $A^T \pi = 0$ for some $\pi \in R^m$, $\pi \geq 0$, $e^T \pi = 1$ where $e = \begin{bmatrix} 1 \\ 1 \\ \vdots \\ \vdots \\ 1 \end{bmatrix}$.

Cox et al.'s (1979) binomial option pricing model is System 2 when System 1 does not hold, i.e., when there is no arbitrage.

## 3. Data and Empirical Results

Daily stocks data of the SSE Index of January 2014 – December 2024 were adopted from the China Stock Market & Accounting Research Database (CSMAR Data). The SSE Composite Index also known as SSE Index is a stock market index of all stocks that are traded at the Shanghai Stock Exchange. Also, daily stocks data of the NYSE Composite Index of January 2014 – December 2024 were adopted from Investing.com and Yahoo Finance. The NYSE Composite Index is a stock market index covering all common stock listed on the New York Stock Exchange. Both the SSE Composite Index and the NYSE Composite Index contain over 2000 stocks.

### 3.1 The Highest p-ratio Stock and the Empirical Efficient Frontier (EEF)

In the case of the SSE Composite stocks, in a certain week, as in eq. (19), we can have:

$$\begin{cases} \text{Shanghai Composite Index (SSEC):} \ \widehat{S_{0SH}} = \frac{1}{1+r}[\pi \cdot S_{0SH}u_{SH} + (1-\pi) \cdot S_{0S} \ d_{SH}] \\ \text{The i} - \text{th stock:} \ S_{0i} = \frac{1}{1+r}[\pi \cdot S_{0i}u_i + (1-\pi) \cdot S_{0i}d_i] \\ \text{Put of the i} - \text{th stock:} \ p_i = \frac{1}{1+r}[\pi \cdot Max(K_i - S_{0i}u_i, 0) + (1-\pi) \cdot Max(K_i - S_{0i}d_i, 0)] \end{cases} \quad (21)$$

and

1. $\widehat{S_{0SH}}$ is the closing price on Monday of Shanghai Composite Index (SSEC), $S_{0SH}u_{SH}$ is the highest closing price of SSEC of the week, $S_{0SH}d_{SH}$ is the lowest closing price of SSEC of the week, and $r$ is equal to the coupon rate (in terms of week) on the 10-year China government bond. (In the case of the NYSE Composite, $r$ is equal to the U.S. 1-year Treasury yield.) Thus, we can calculate $\pi$ and $1 - \pi$. If it is found that $1 > \pi > 0$ does not hold, then use this week's and the previous week's closing prices (and $r$ becomes $2r$) to calculate $\pi$.

2. For the i-th company, $S_{0i}u_i$ is the highest closing price of the i-th stock of the week, and $S_{0i}d_i$ is the lowest closing price of the i-th stock of the week.[4] Let $K_i = (1 + \delta)S_{0i}$ and $\delta = r$. With $K_i$, $r$, $\pi$, $S_{0i}u_i$ and $S_{0i}d_i$, we can estimate $p_i$, the i-th stock's put option closing price on Monday of the week. The p-index of the i-th stock in this week is: $\frac{p_i}{K_i} = \frac{p_i}{(1+r)S_{0i}}$ , which is the insurance fee for each insured dollar so that the i-th stock can achieve at least $r$ rate of return at the end of the week. If it is found that $p_i > 0$ does not hold, then use this week's and the previous week's closing prices (and $r$ becomes $2r$) to calculate $p_i$.

---

[4] We have considered forward adjustment in stock prices. For example, suppose Stock A was priced at $10/share. Due to an ex-rights event, it becomes $8/share. Selecting forward adjustment would display both the current and previous day's price as $8/share. All earlier prices would be proportionally scaled down.

We have done one-week and one-month holding periods investments in both SSE Composite and NYSE Composite stocks. In the case of monthly investment in SSE Composite stocks, a 2-month cycle serves as an example.

(i) As shown in Figure 1, we first use the closing prices of stocks of Month 1 (i.e., October, 2024) to calculate the p-indexes (i.e., risk levels) of the stocks and then, use equations (17) and (18) to construct the empirical efficient frontier (EEF) over the period. The empirical efficient frontier (EEF), the set of all possible portfolios that offer the best return for a given level of risk, is a piecewise linear curve that has non-negative slope at every point. Also, as shown in Figure 2, the highest p-ratio stock (i.e., eq. (7)) is the tangent stock on the EEF where a line drawn from the risk-free rate is tangent to the curve. The risk-free asset has a zero p-index.

(ii) In Month 2, three investment strategies are adopted: (a) H-p-ratio: the highest p-ratio stock; (b) EEF stocks: equally weighted portfolio of stocks on the EEF; and (c) H-p-ratio & risk-free-rate: equally weighted portfolio of the highest p-ratio stock and the risk free asset. Also, stocks are bought at the closing price on the first trading date and sold at the closing price on the last trading date.

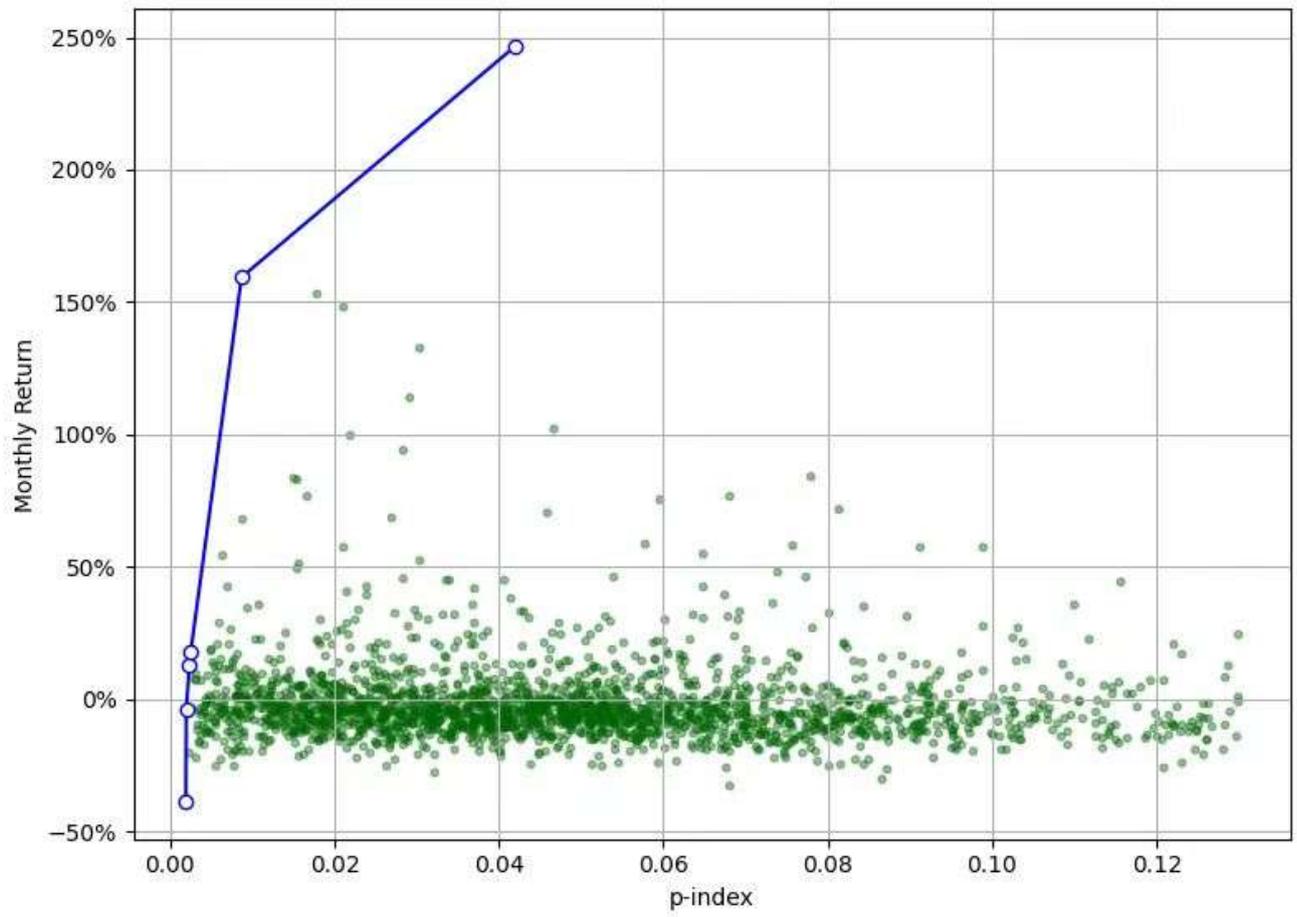

Figure 1 The empirical efficient frontier (EEF) in China's SSE (October 2024)

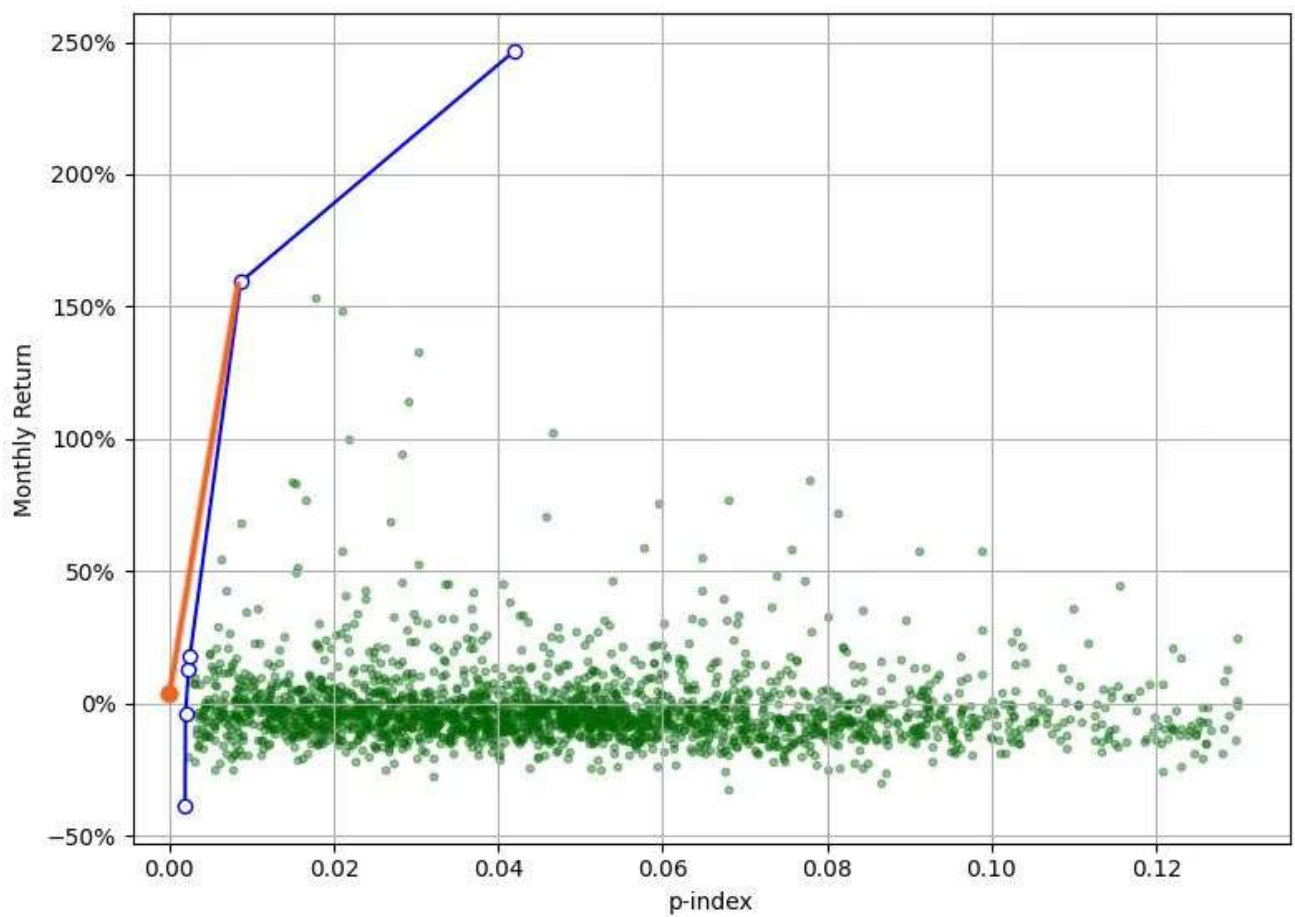

Figure 2 The empirical efficient frontier (EEF) and highest p-ratio stock in China's SSE (October 2024)

1. Reinvesting: For weekly (or monthly) strategy: Invest $1 to buy stocks at the closing price on the first trading day, sell them at the closing price on the last trading day, and then reinvest the proceeds into other stocks at the start of the subsequent week (or month).

As Table 1 indicates, with one-week holding period, all three strategies based on SSE Composite Index stocks generate exceptionally high returns, with the highest p-ratio strategy producing the largest annualized rate of return, at 499.97%. However, with one-month holding period, all three strategies based on SSE Composite Index stocks deliver negative returns. The three strategies with both one-week and one-month periods on NYSE Composite Index stocks also generate negative returns. This indicates that the reinvesting strategy may not be suitable for NYSE Composite Index stocks.

Table 1 Returns of reinvesting strategies (2014-2024)

| Strategies | SSE | | NYSE | |
| --- | --- | --- | --- | --- |
| | Weekly | Monthly | Weekly | Monthly |
| H-p-ratio | 36259664737.39% (Annualized: 499.97%) | -96.94% (Annualized: -27.17%) | -99.88% (Annualized: -45.74%) | -99.59% (Annualized: -39.33%) |
| EEF stocks | 128793211.43% (Annualized: 259.29%) | -16.56% (Annualized: -1.63%) | -45.66% (Annualized: -5.39%) | -91.56% (Annualized: -20.12%) |
| H-p-ratio & risk-free-rate | 208440133.18% (Annualized: 275.36%) | -41.23% (Annualized: -4.72%) | -73.90% (Annualized: -11.50%) | -82.79% (Annualized: -14.78%) |

2. Non-reinvesting: For weekly (or monthly) strategy: In each week (or month) invest $1 to buy stocks at the closing price on the first trading day, sell them at the closing price on the last trading day.

As shown in Table 2, for SSE Composite Index stocks, the highest p-ratio strategy with one-week holding period still yields the highest annualized rate of return (36.48%), followed by the EEF stocks strategy (29.98%). For NYSE Composite Index stocks, the one-week EEF stocks strategy produces a 10.44% annualized return. It seems that one-week holding period outperforms one-month holding period for both exchanges. Furthermore, Figures 3 and 4 demonstrate that the one-week EEF stocks strategy exhibits smoother performance for SSE stocks than for NYSE stocks.

Table 2 Returns of non-reinvesting strategies (2014-2024)

| Strategies | SSE | | NYSE | |
| --- | --- | --- | --- | --- |
| | Weekly | Monthly | Weekly | Monthly |
| H-p-ratio | 2961.42% (Annualized: 36.48%) | 125.36% (Annualized: 7.67%) | -51.61% (Annualized: -4.69%) | -154.31% (Annualized: -14.03%) |
| EEF stocks | 1689.76% (Annualized: 29.98%) | 166.12% (Annualized: 9.31%) | 198.13% (Annualized: 10.44%) | -127.82% (Annualized: -11.62%) |

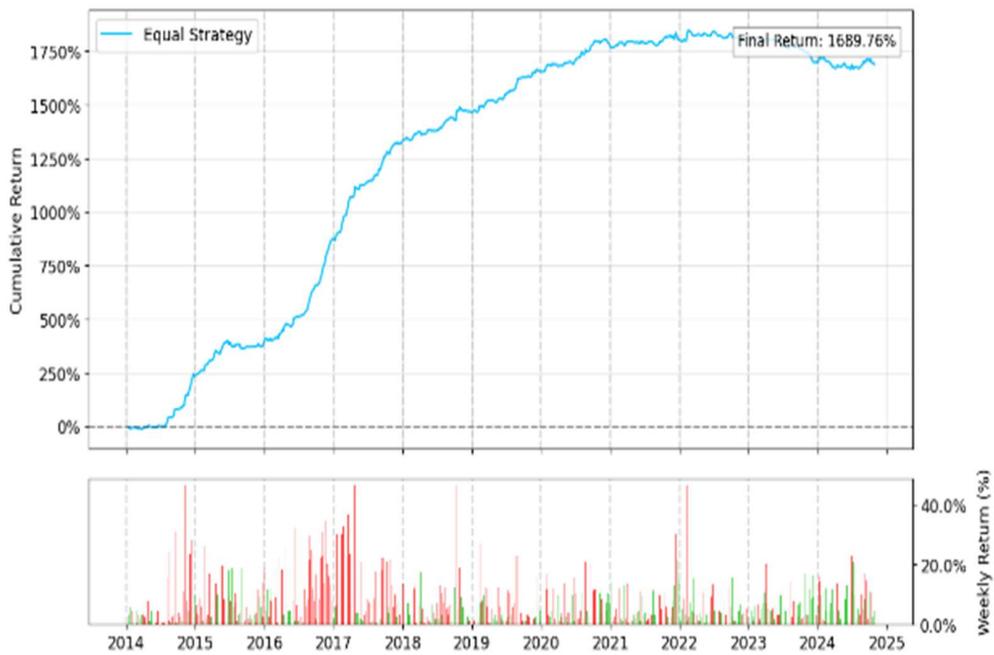

Figure 3 Returns of non-reinvesting weekly EEF stocks (equally-weighted) strategy for SSE

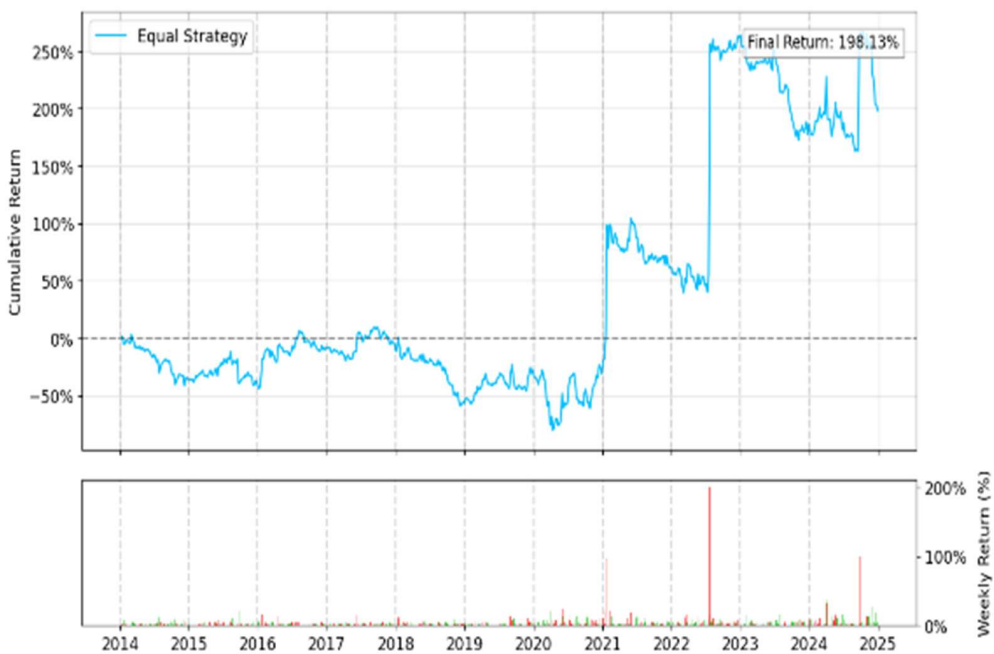

Figure 4 Returns of non-reinvesting weekly EEF stocks (equally-weighted) strategy for NYSE

3. Adjustments for SSE Composite Index stocks

Due to price limit rules on SSE Composite Index stocks, some stocks may hit limit up as soon as the market opens and become unavailable for purchase. To tackle this problem, the following rules are employed to adjust the returns of stocks listed on the SSE Composite Index in Tables 1 and 2.

a. Examine whether the lowest price falls below the limit-up price (limit-up price = previous day's closing price × individual stock price limit). If the day's lowest price equals the limit-up price, defer the buy order to the next trading day.
b. If the stock is purchasable during the day, identify the lowest intraday price reached (target entry level). Suppose that stock A opens at ¥10.00 (with a 10% price limit). If its lowest price reaches ¥10.80 or lower (i.e., $\leq 8\%$ gain), execute the buy at this price.
c. If the stock is purchasable but condition b is unmet (e.g., the price fluctuates strictly between ¥10.80 and ¥11.00 without dipping to/below ¥10.80), buy at the closing price.

As shown in Table 3, with this conservative approach the reinvest EEF stocks strategy with one-week holding period yields the highest annualized rate of return (22.52%), followed by the non-reinvest EEF stocks strategy (15.27%). The reinvest highest p-ratio strategy with one-week holding period suffers loss, but the non-reinvest highest p-ratio strategy with one-week holding period yields a positive annualized rate of return (13.31%).

Table 3 Adjusted returns of returns of stocks listed on the SSE Composite Index (2014-2024)

| Strategies | Reinvest | | Non-reinvest | |
|---|---|---|---|---|
| | Weekly | Monthly | Weekly | Monthly |
| H-p-ratio | -81.69% (Annualized: -13.54% | -99.71% (Annualized: -41.21% | 295.45% (Annualized: 13.31%) | -186.41% (Annualized: -16.94% |
| EEF stocks | 934.24% (Annualized: 22.52%) | -79.83% (Annualized: -13.54% | 377.59% (Annualized: 15.27%) | -31.07% (Annualized: -2.82% |

4. Left frontier (EEF stocks) and right frontier investments

Using equation (18), we derive the curve of maximum rate of return portfolios. The non-negative slope segment (the left frontier: the empirical efficient frontier in Figure 2) represents portfolios with high returns and a low p-index, while the non-positive slope segment (the right frontier) represents portfolios with high returns and a high p-index. A key question is whether stocks on the right frontier (characterized by high return and high p-index) in one period generate high returns in the subsequent period. Table 4 shows that under the weekly EEF stocks strategy, for NYSE Composite Index stocks, the right frontier yields a higher annualized return (14.41%). Conversely, for SSE Composite Index stocks, the left frontier yields a higher annualized return than the right frontier (weekly: 29.98% vs. 23.34%; monthly: 9.31% vs. 8.11%). This may indicate that high-return stocks in the SSE Composite Index are not consistently associated with high risk (as proxied by the p-index).

Table 4 Returns of non-reinvesting left (EEF) and right frontiers strategies (2014-2024)

|  | SSE | | NYSE | |
|---|---|---|---|---|
|  | Left frontier | Right frontier | Left frontier | Right frontier |
| Weekly | 1689.76% (Annualized: 29.98%) | 905.21% (Annualized: 23.34%) | 198.13% (Annualized: 10.44%) | 339.40% (Annualized: 14.41%) |
| Monthly | 166.12% (Annualized: 9.31%) | 135.69% (Annualized: 8.11%) | -127.82% (Annualized: -11.62%) | -76.81% (Annualized: -12.44%) |

## 3.2 The p-index and Multifactor Models

We now examine the relationship between monthly return and p-index (risk) for NYSE and SSE Composite Index stocks during 2014-2024. First, in each month, after calculating the p-index for each stock, we rank these p-indexes from high to low and divide them into ten groups (ten equally weighted portfolios). Second, run the following cross-sectional regression for 132 months:

$$R_{pt} = \gamma_{0t} + \gamma_{1t} v_{p,t-1} + \gamma_{2t} v_{p,t-1}^2 + \varepsilon_{pt} \ , \ p = 1, 2, \ldots 10, \quad (22)$$

where $R_{pt}$ is the t-th month return of portfolio $p$; $v_{p,t-1}$ is the average of the $v_i$ (the p-index) for stocks in portfolio $p$ in the $t-1$ month; and $v_{p,t-1}^2$ is the average of the squared values of these $v_i$ in the

$t-1$ month; and $\varepsilon_{pt}$ is the error term. Newey and West (1987) corrected t-statistics (using 4 lags) are used.

For NYSE Composite Index stocks (Table 5), a positive linear relationship exists between return and the p-index, with $\gamma_{1t}$ significant at the 0.10 level. No such relationship is evident for the SSE Composite Index stocks (Table 6), as both $\gamma_{1t}$ and $\gamma_{2t}$ are statistically insignificant (p > 0.10).

Table 5 Average values of the estimated coefficients of the two-parameter model (NYSE)

| Coefficient | Average | p value | t statistics |
|---|---|---|---|
| $\gamma_{0t}$ | 0.0014 | 0.1090 | 1.2376 |
| $\gamma_{1t}$ | 19.1551* | 0.0751* | 1.4473* |
| $\gamma_{2t}$ | -10201.3277 | 0.2159 | -0.7887 |

* Significant at the 0.10 level.

Table 6 Average values of the estimated coefficients of the two-parameter model (SSE)

| Coefficient | Average | p value | t statistics |
|---|---|---|---|
| $\gamma_{0t}$ | 0.0125* | 0.0575* | 1.8993* |
| $\gamma_{1t}$ | -0.0038 | 0.1388 | -1.4803 |
| $\gamma_{2t}$ | 0.0000 | 0.1517 | 1.4334 |

* Significant at the 0.10 level.

During 2014-2024, in each month, rank stocks' p-indexes from high to low and divide them into ten equally weighted portfolios, and then calculate each portfolio's return in the next month. Also, for each portfolio we can run Fama-French's five-factor model (FF5) center on the time-series regression,

$$R_{pt}^e = a_p + \beta_{1p}MKT\_RF_t + \beta_{2p}SMB_t + \beta_{3p}HML_t + \beta_{4p}RMW_t + \beta_{5p}CMA_t + u_{it} , \qquad (23)$$

where $R_{pt}^e = R_{pt} - R_{Ft}$ is the monthly return on the considered portfolio $p$ minus the monthly risk-free rate; $a_p$ (FF5 Alpha) is the portfolio's excess return not explained by the five factors; $MKT\_RF_t$ is equity market return in excess of the risk-free rate; $SMB_t$ is Small Minus Big, return spread between small cap and large cap stocks; $HML_t$ is High Minus Low, return spread between high book-to-market and low book-to-market stocks; $RMW_t$ Robust Minus Weak, return spread between high and low profitability stocks; $CMA_t$ is Conservative Minus Aggressive, return spread between stocks with low and high investment ratios; and $u_{it}$ is the error term. Five factors of NYSE Composite Index stocks were adopted from Kenneth French's data library, and five factors of SSE Composite Index stocks were adopted from CSMAR.

For NYSE Composite Index stocks (Table 7), with the reinvesting strategy, the highest p-index portfolio (H) yields the largest annualized return (20.58%), while the lowest p-index portfolio (L) yields the lowest (−1.51%). As shown in Figure 5, the H-L portfolio (long H, short L) generates an annualized return of 22.85%. Using eq. (23), in Table 7, FF5 Alpha (the intercept term) for portfolio L is -0.28%, and for portfolio H, it is 0.93%; both are significant at the 0.10 level.

Table 7 Performance of different p-index portfolios (NYSE)

| P-index portfolio | Annualized return | Annualized volatility | t stat of returns | FF5 Alpha avg. | t stat of FF5 Alpha |
|---|---|---|---|---|---|
| L | -1.51% | 7.49% | -0.5362 | -0.28% | -1.5750 |
| 2 | 1.25% | 6.48% | 0.7440 | -0.17% | -1.4693 |
| 3 | 2.45% | 7.86% | 1.1475 | -0.16% | -1.2720 |
| 4 | 3.73% | 9.39% | 1.4466 | -0.13% | -1.0267 |
| 5 | 3.99% | 11.37% | 1.3295 | -0.20% | -1.4121 |
| 6 | 5.53% | 13.49% | 1.5488 | -0.14% | -0.9086 |
| 7 | 6.60% | 16.36% | 1.5727 | -0.13% | -0.7697 |
| 8 | 9.13% | 19.54% | 1.8142 | 0.01% | 0.0323 |
| 9 | 11.01% | 24.81% | 1.8237 | 0.12% | 0.4797 |
| H | 20.58% | 48.21% | 2.0578 | 0.93% | 1.4045 |

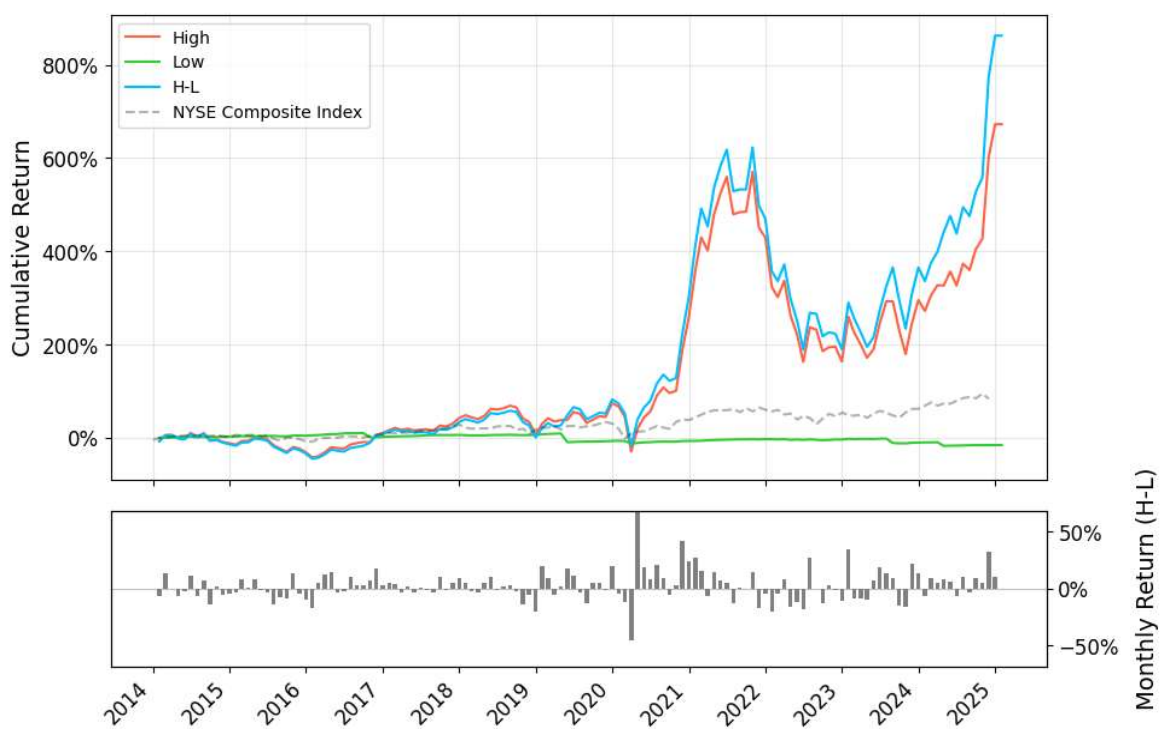

Figure 5 Returns of high, low and H-L p-index portfolios (NYSE)

For SSE Composite Index stocks (Table 8), the lowest p-index portfolio (L) yields the largest annualized return (27.36%), while the highest p-index portfolio (H) yields an annualized return of 9.22%. As shown in Figure 6, the L-H portfolio (long L, short H) generates an annualized return of 14.41%. Using eq. (23), in Table 8, FF5 Alpha (the intercept) for portfolio L is 0.0239, and for portfolio H, it is 0.0128; both are significant at the 0.10 level.

Table 8 Performance of different p-index portfolios (SSE)

| P-index portfolio | Annualized return | Annualized volatility | t stat of returns | FF5 Alpha avg. | t stat of FF5 Alpha |
|---|---|---|---|---|---|
| L | 27.36% | 29.98% | 3.1277 | 0.0239 | 3.3560 |
| 2 | 9.20% | 25.55% | 1.5654 | 0.0109 | 1.6919 |
| 3 | 7.91% | 25.59% | 1.4101 | 0.0100 | 1.5272 |
| 4 | 6.78% | 26.10% | 1.2646 | 0.0092 | 1.3964 |
| 5 | 9.81% | 25.64% | 1.6309 | 0.0115 | 1.7639 |
| 6 | 11.87% | 27.99% | 1.7779 | 0.0144 | 2.0705 |
| 7 | 11.48% | 26.02% | 1.8132 | 0.0132 | 2.0144 |
| 8 | 7.97% | 26.57% | 1.3937 | 0.0111 | 1.6581 |
| 9 | 9.41% | 26.69% | 1.5520 | 0.0125 | 1.8673 |
| H | 9.22% | 29.85% | 1.4627 | 0.0128 | 1.7442 |

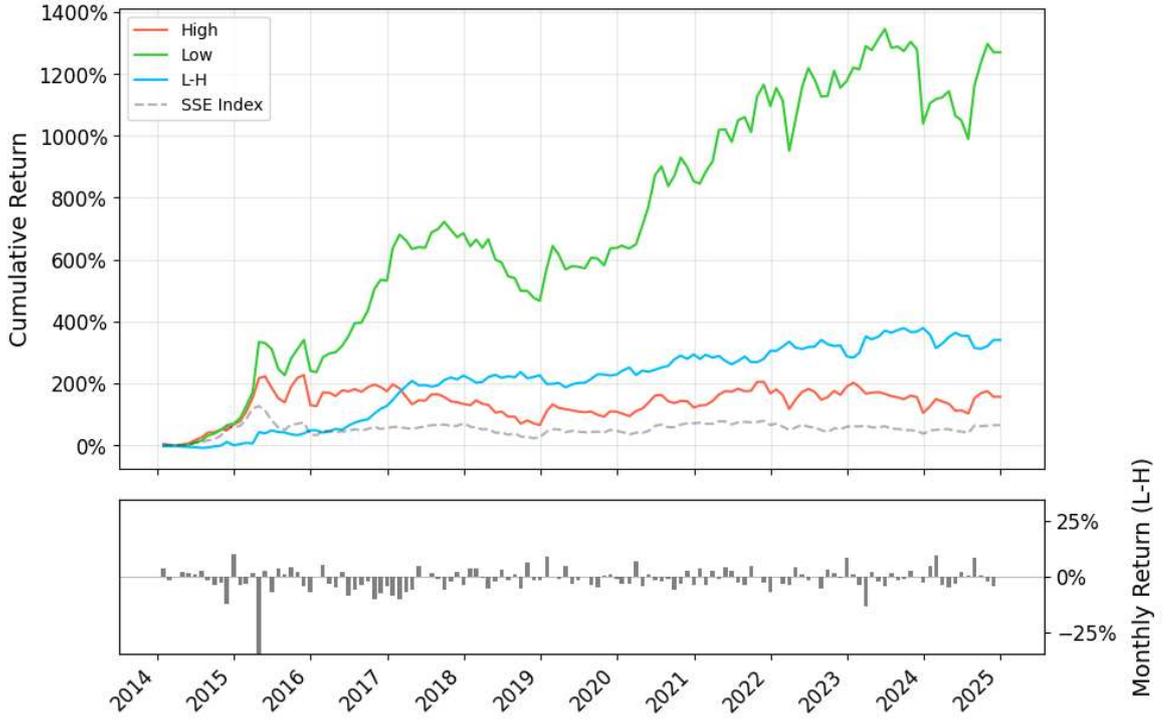

Figure 6 Returns of high, low and L-H p-index portfolios (SSE)

As a first step in examining the relationship between the p-index and the five common factors, we estimate the following time-series regression for all stocks $i = 1,..,n$:

$$R_{it}^e = a_{it} + \beta_{vit}v_{it} + \beta_{1it}MKT\_RF_t + \beta_{2it}SMB_t + \beta_{3it}HML_t + \beta_{4it}RMW_t + \beta_{5it}CMA_t + u_{it} \ , (24)$$

where $v_{it}$ is the i-th stock's p-index in the t-th month, $\beta_{vit}$ is the factor exposure of the p-index, and $u_{it}$ is the error term. Next, we run the following cross-sectional regression for 132 months ($t = 1,..,132$),

$$R_{it}^e = \alpha_{it} + \beta_{vi,t-1}\lambda_{vit} + \beta_{1i,t-1}\lambda_{1it} + \beta_{2i,t-1}\lambda_{2it} + \beta_{3i,t-1}\lambda_{3it} + \beta_{4i,t-1}\lambda_{4it} \\ + \beta_{5i,t-1}\lambda_{5i} + \eta_{it} \ , \qquad (25)$$

where $\lambda_{vit}$ is the factor return of the p-index factor, $\lambda_{jit}$ is the factor return of the j-th common factor ($j = 1,..,5$), and $\eta_{it}$ is the error term.

Table 9 shows that for NYSE Composite Index stocks, $\lambda_{vit}$ (the factor return of the p-index factor) is significant at the 0.10 level, and $\lambda_{2it}$ (the factor return of SMB) is positive and significant at the 0.05 level. This indicates that the p-index can provide incremental information. Table 10 shows that without incorporating the p-index, the traditional five-factor model performs poorly (all the coefficients are insignificant at the 0.10 level).

Table 9 P-index and five-factor model (NYSE)

| Factor  | λ avg.    | t-stat   | p value  |
|---------|-----------|----------|----------|
| P-index | -0.0012*  | -1.8098* | 0.0746*  |
| MKT_RF  | 0.6412    | 1.3821   | 0.1713   |
| SMB     | 0.5359**  | 2.0572** | 0.0434** |
| HML     | 0.5651    | 1.3455   | 0.1828   |
| RMW     | -0.0336   | -0.2399  | 0.8111   |
| CMA     | 0.0395    | -0.1566  | 0.8761   |

* Significant at the 0.10 level. ** Significant at the 0.05 level.

Table 10 Five-factor model (NYSE)

| Factor | λ avg.  | t-stat  | p value |
|--------|---------|---------|---------|
| MKT_RF | -0.5677 | -1.0553 | 0.2950  |
| SMB    | -0.3250 | -1.1597 | 0.2502  |
| HML    | -0.2518 | -0.8075 | 0.4221  |
| RMW    | -0.2553 | -1.5537 | 0.1248  |
| CMA    | -0.2119 | -0.9090 | 0.3665  |

For SSE Composite Index stocks (Table 11), $\lambda_{vit}$, the factor return of the p-index, is insignificant at the 0.10 level (p value = 0.1018), but the factor returns of SMB and RMW are significant at least at the 0.10 level. Table 12 shows that with the five-factor model, the factor return of SMB is still significant at the 0.10 level.

Table 11 P-index and five-factor model (SSE)

| Factor  | λ avg.  | t-stat    | p value   |
|---------|---------|-----------|-----------|
| P-index | 0.0024  | 1.6363    | 0.1018    |
| MKT_RF  | 0.0012  | 0.4239    | 0.6716    |
| SMB     | -0.0048 | -1.9837** | 0.0473**  |
| HML     | 0.0015  | 0.8071    | 0.4196    |
| RMW     | 0.0030  | 1.8483*   | 0.0646*   |
| CMA     | -0.0006 | -0.3781   | 0.7053    |

* Significant at the 0.10 level. ** Significant at the 0.05 level.

Table 12 Five-factor model (SSE)

| Factor | λ avg. | t-stat | p value |
|---|---|---|---|
| MKT_RF | 0.0007 | 0.2406 | 0.8099 |
| SMB | -0.0047* | -1.8065* | 0.0708* |
| HML | 0.0017 | 0.8177 | 0.4135 |
| RMW | 0.0028 | 1.6001 | 0.1096 |
| CMA | -0.0002 | -0.1200 | 0.9045 |

* Significant at the 0.10 level.

We now use the p-index to construct the modified p-ratio:

$$Modified\ p-ratio_{it} = \frac{[R_{it}-\min_j(R_{jt})]/[\max_j(R_{jt})-\min_j(R_{jt})]}{\{[v_{it}-\min_j(v_{jt})]/[\max_j(v_{jt})-\min_j(v_{jt})]\}+\varepsilon} \qquad (26)$$

where $R_{it}$ is the t-th monthly return of the i-th stock, $v_{it}$ is the i-th stock's p-index in the t-th month, and $\varepsilon = 10^{-4}$. In equations (24)-(25) and Tables 7-12, substitute the p-index ($v_{it}$) with the modified p-ratio, we have the following results:

1. For NYSE Composite Index stocks (Table 13), the lowest p-ratio portfolio (L) yields the highest annualized rate of return (23.78%), while the largest p-ratio portfolio (H) yields the lowest (-1.48%). As shown in Figure 7, the L-H portfolio (long L, short H) generates an annualized return of 25.9%.

2. For SSE Composite Index stocks (Table 14), the lowest p-ratio portfolio (L) yields the highest annualized rate of return (11.87%), while the largest p-ratio portfolio (H) yields the lowest (3.83%). As shown in Figure 8, the L-H portfolio (long L, short H) generates an annualized return of 6.11%.

3. For NYSE Composite Index stocks (Table 15), $\lambda_{vi}$ (the factor return of the p-ratio factor) is significant at the 0.10 level, but the factor returns of Fama-French's five factors are not. This indicates that the p-ratio can provide incremental information for NYSE Composite Index stocks. Table 16 shows that for SSE Composite Index stocks, all factor returns are insignificant at the 0.10 level.

Table 13 Performance of different p-ratio portfolios (NYSE)

| P-ratio portfolio | Annualized return | Annualized volatility | t stat of returns | FF5 Alpha avg. | t stat of FF5 Alpha |
|---|---|---|---|---|---|
| L | 23.78% | 46.02% | 2.2714 | 1.15% | 1.7962 |
| 2 | 9.73% | 25.27% | 1.6535 | -0.03% | -0.1013 |
| 3 | 9.26% | 19.86% | 1.8190 | -0.01% | -0.0479 |
| 4 | 6.94% | 16.56% | 1.6265 | -0.14% | -0.7984 |
| 5 | 5.33% | 13.99% | 1.4669 | -0.19% | -1.2177 |
| 6 | 4.40% | 11.38% | 1.4432 | -0.18% | -1.3250 |
| 7 | 3.87% | 9.36% | 1.4985 | -0.16% | -1.2306 |
| 8 | 2.50% | 7.91% | 1.1643 | -0.18% | -1.4319 |
| 9 | 1.64% | 6.16% | 0.9759 | -0.14% | -1.3524 |
| H | -1.48% | 7.19% | -0.5536 | -0.29% | -1.7120 |

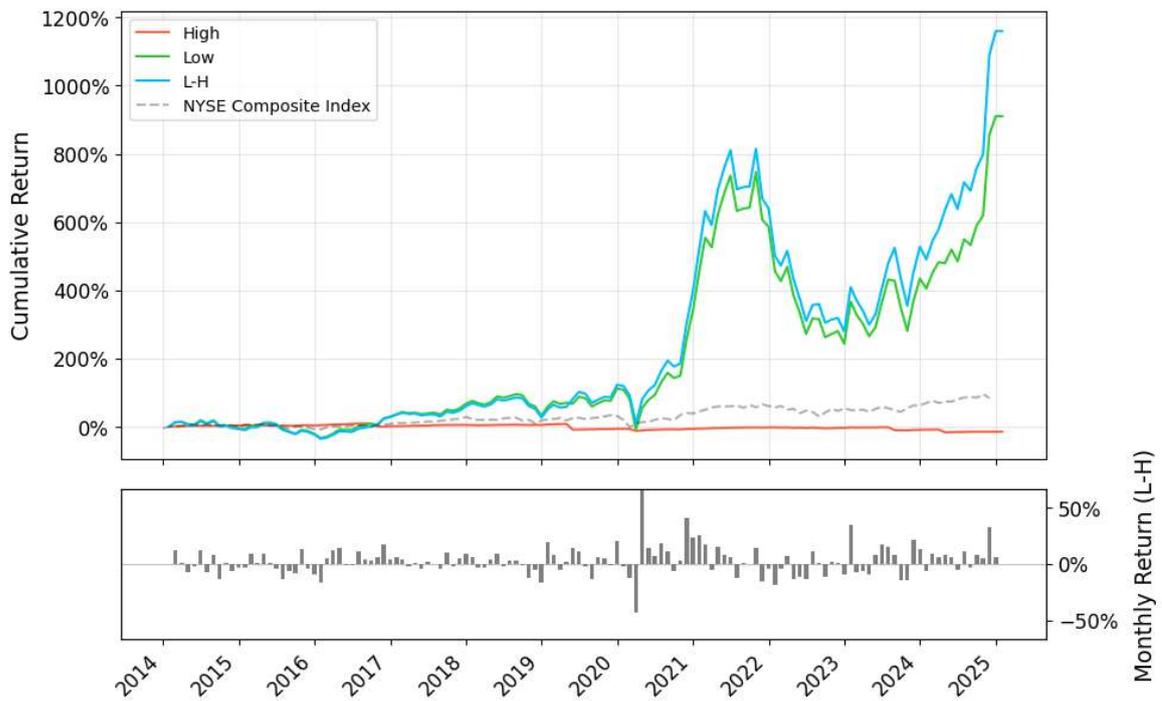

Figure 7 Returns of high, low and L-H p-ratio portfolios (NYSE)

Table 14 Performance of different p-ratio portfolios (SSE)

| P-ratio portfolio | Annualized return | Annualized volatility | t stat of returns | FF5 Alpha avg. | t stat of FF5 Alpha |
|---|---|---|---|---|---|
| L | 11.87% | 29.57% | 1.7352 | 0.0148 | 2.0091 |
| 2 | 10.41% | 27.44% | 1.6454 | 0.0133 | 1.9249 |
| 3 | 9.61% | 26.56% | 1.5800 | 0.0121 | 1.8094 |
| 4 | 12.97% | 26.87% | 1.9396 | 0.0146 | 2.1689 |
| 5 | 10.81% | 25.76% | 1.7420 | 0.0123 | 1.9004 |
| 6 | 11.36% | 26.02% | 1.7986 | 0.0128 | 1.9459 |
| 7 | 7.29% | 26.25% | 1.3233 | 0.0091 | 1.3634 |
| 8 | 6.41% | 25.75% | 1.2277 | 0.0089 | 1.3660 |
| 9 | 4.80% | 25.80% | 1.0336 | 0.0074 | 1.1333 |
| H | 3.83% | 26.99% | 0.9145 | 0.0064 | 0.9451 |

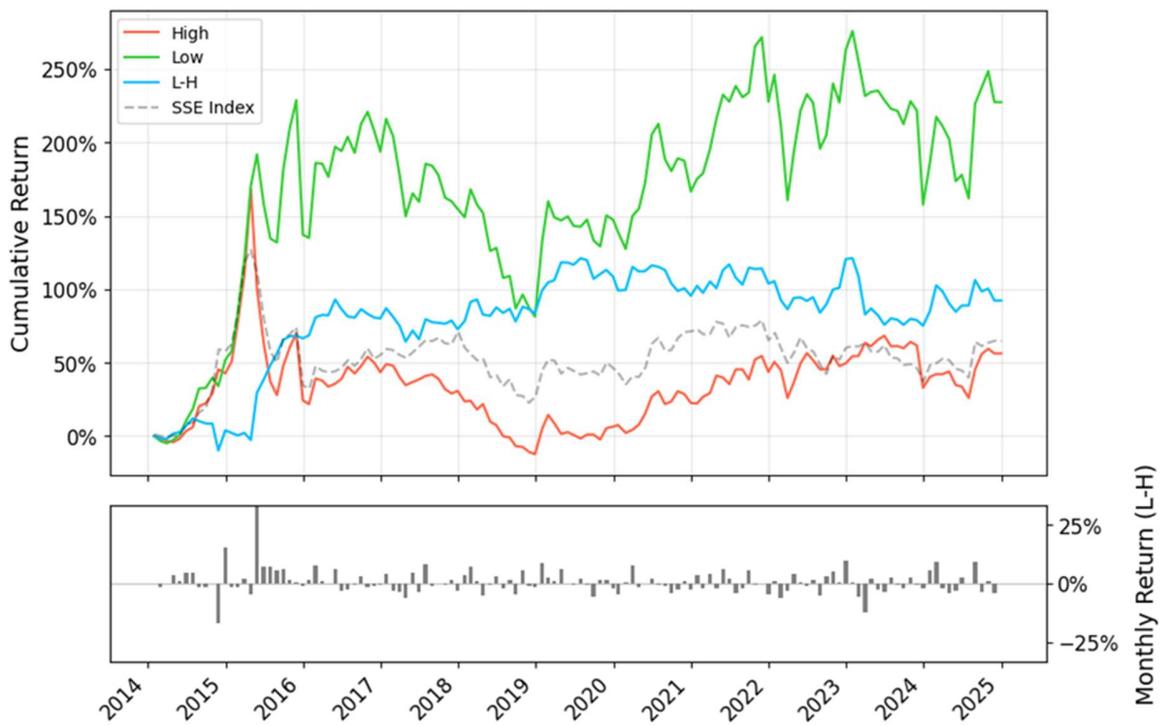

Figure 8 Returns of high, low and L-H p-ratio portfolios (SSE)

Table 15 P-ratio and five-factor model (NYSE)

| Factor | λ avg. | t-stat | p value |
|---|---|---|---|
| P-ratio | 0.7214* | 1.9478* | 0.0514* |
| MKT_RF | -0.5539 | -0.7560 | 0.4496 |
| SMB | -0.1358 | -0.4321 | 0.6657 |
| HML | -0.1226 | -0.4838 | 0.6285 |
| RMW | -0.1480 | -0.8208 | 0.4118 |
| CMA | -0.1141 | -0.5948 | 0.5520 |

* Significant at the 0.10 level. ** Significant at the 0.05 level.

Table 16 P-ratio and five-factor model (SSE)

| Factor | λ avg. | t-stat | p value |
|---|---|---|---|
| P-ratio | -0.0039 | -0.0148 | 0.9882 |
| MKT_RF | 0.0034 | 1.1608 | 0.2457 |
| SMB | 0.0014 | 0.4334 | 0.6647 |
| HML | 0.0000 | -0.0011 | 0.9991 |
| RMW | -0.0013 | -0.6662 | 0.5053 |
| CMA | 0.0014 | 0.7123 | 0.4763 |

## 4. Concluding Remarks

Risk level can be defined as the likelihood that a party will fail to meet a stated obligation or promise. This paper has used European put option to construct the p-index to measure the underlying asset's risk level. The p-index measures the insurance fee for each insured dollar to guarantee that the asset achieves at least a $\delta$ rate of return on a specified future date. We use the p-index to construct both the p-ratio and the empirical efficient frontier (EEF) to examine the performance of different investment strategies for stocks listed on both China's SSE Composite Index and the US NYSE Composite Index from 2014 to 2024.

The empirical results have shown:
1. For the three investment strategies (the highest p-ratio stock; equally weighted portfolio of stocks on the EEF; and equally weighted portfolio of the highest p-ratio stock and the risk-free asset), with one-week holding period and reinvesting, all three strategies based on SSE Composite Index stocks generate exceptionally high returns, with the highest p-ratio strategy producing the highest annualized rate of return, at 499.97%. However, with one-month holding period, all three strategies based on SSE Composite Index stocks deliver negative returns. The three strategies with both one-week and one-month periods on NYSE Composite Index stocks also generate negative returns.

2. With non-reinvesting, for SSE Composite Index stocks, the highest p-ratio strategy with one-week holding period yields the highest annualized rate of return (36.48%), followed by the EEF stocks strategy (29.98%). For NYSE Composite stocks, the one-week EEF strategy produces a 10.44% annualized return. It seems that one-week holding period outperforms one-month holding period for both exchanges.

3. After considering the effect of the price limit rules on SSE Composite Index stocks, it is found that the reinvest EEF stocks strategy with one-week holding period yields the highest annualized rate of return (22.52%), followed by the non-reinvest EEF stocks strategy (15.27%). The reinvest highest p-ratio strategy with one-week holding period suffers loss, but the non-reinvest highest p-ratio strategy with one-week holding period yields a positive annualized rate of return (13.31%).

4. Under the one-week EEF strategy, for NYSE Composite Index stocks, the right frontier yields a higher annualized return (14.41%). Conversely, for SSE Composite Index stocks, the left frontier (stocks on the empirical efficient frontier) yields a higher annualized return than the right frontier (weekly: 29.98% vs. 23.34%; monthly: 9.31% vs. 8.11%). This suggests that for SSE Composite Index stocks, higher returns are not systematically associated with greater risk when proxied by the p-index.

5. For NYSE Composite Index stocks, there is a positive linear relationship between monthly return and the p-index. No such relationship is evident for SSE Composite Index stocks.

6. For NYSE Composite Index stocks, with the reinvesting strategy, the highest p-index portfolio (H) yields the largest annualized return (20.58%), while the lowest p-index portfolio (L) yields the lowest (−1.51%). The H-L portfolio (long H, short L) generates an annualized return of 22.85%. For SSE Composite Index stocks, the lowest p-index portfolio (L) yields the largest annualized return (27.36%), while the highest p-index portfolio (H) yields an annualized return of 9.22%. The L-H portfolio (long L, short H) generates an annualized return of 14.41%.

7. The traditional five-factor model performs poorly for NYSE Composite Index stocks. After adding the p-index as the sixth factor, it is found that the p-index can provide incremental information. The results of using the five-factor model for SSE Composite Index stocks show that the factor return of SMB is significant at the 0.10 level. After adding the p-index as the sixth factor, it is found that the p-index may not provide incremental information, and the factor returns of SMB and RMW are significant at least at the 0.10 level. Also, including the p-ratio as the sixth factor in place of the p-index within the five-factor model yields almost the same results.